\documentclass[graybox,natbib,nosecnum]{svmult}
\bibpunct{(}{)}{;}{a}{}{,} 

\pdfoutput=1   

\usepackage{mathptmx}       
\usepackage{helvet}         
\usepackage{courier}        
\usepackage{type1cm}        

\usepackage{makeidx}         
\usepackage{graphicx}        
\usepackage{multicol}        
\usepackage[bottom]{footmisc}
\usepackage[normalem]{ulem}	
\usepackage{hyperref}  

\usepackage{soul}   

\usepackage{amsmath}
\usepackage{amssymb}
\hypersetup{
   colorlinks=true,
   citecolor=blue,
   linkcolor=blue,
   urlcolor=black
   }

\makeindex             

\begin{document}

\title*{Exoplanet phase curves: observations and theory}
\author{Vivien Parmentier and Ian J. M. Crossfield}
\institute{Vivien Parmentier \at University of Arizona, Lunar and Planetary Laboratory, 1629 E University Blvd, Tucson, AZ, USA\email{vivien@lpl.arizona.edu}
\and Ian J. M. Crossfield \at Department of Physics, Massachusetts Institute of Technology, Cambridge, MA, USA  \email{iancross@mit.edu}}
%
%
\maketitle

\abstract{Phase curves are the best technique to probe the {three dimensional structure} of exoplanets' atmospheres. In this chapter we first review current exoplanets phase curve observations and the particular challenges they face. We then describe the different physical mechanisms shaping the atmospheric phase curves of highly irradiated tidally locked exoplanets. Finally, we discuss the potential for future missions to further advance our understanding of these new worlds. 
}

\section{Observing exoplanetary phase curves}

\subsection{Origin and shape of a phase curve}
Planetary atmospheres are intrinsically three-dimensional objects, with both small- and large-scale variations of temperature, chemistry, and cloud coverage. This is even more important for the current population of characterizable exo-atmospheres, as most of them belong to planets in close-in orbits, probably tidally locked, with a large day/night temperature contrast induced by permanent radiative forcing. When ignored, the large spatial inhomogeneities of these planets can lead to a biased interpretation of transiting and secondary eclipse observations~\citep{Line2016,Feng2016}. 

Observing the {phase curve} of an exoplanet (i.e., the time-dependent change in the brightness of a planet as seen from Earth during one orbital period) is the most straightforward way to probe the planet's longitudinal structure. The brightness of the planet is determined by the combined emitted and reflected light in the particular bandpass of the observations. For transiting planets, the shape and amplitude of the phase curve are determined by longitudinal inhomogeneities. Inhomogeneous illumination is always present, as over the course of one orbit, we see different hemispheres of the planet, ranging from its {dayside} (before and after the eclipse of the planet by the star) to its {nightside}. Inhomogeneities in temperatures can be probed by the thermal emission of the planet, with a nightside that is usually colder and thus dimmer than the dayside. Finally, inhomogeneous chemical composition and cloud coverage can be determined through the phase curve's wavelength dependence and shape.

Short-period planets are good targets for atmospheric characterization through phase curve observations: their short orbit makes it easier to monitor the system during a full orbit of the planet. Moreover, planets in orbits  $\lesssim$10 days are likely tidally locked~\citep{Guillot1996,Parmentier2015a}, meaning that their rotation period is their orbital period and that the longitude of the Earth-facing hemisphere can be related to the phase of the orbit. Finally, tidally locked planets in short orbit have a short radiative timescale, a weak Coriolis force, and an input of energy at large scale, making them likely to have planetary-scale atmospheric features that can be observed in the hemispherically averaged planetary flux~\citep{Showman2002}.
\begin{figure}
\includegraphics[width=\linewidth]{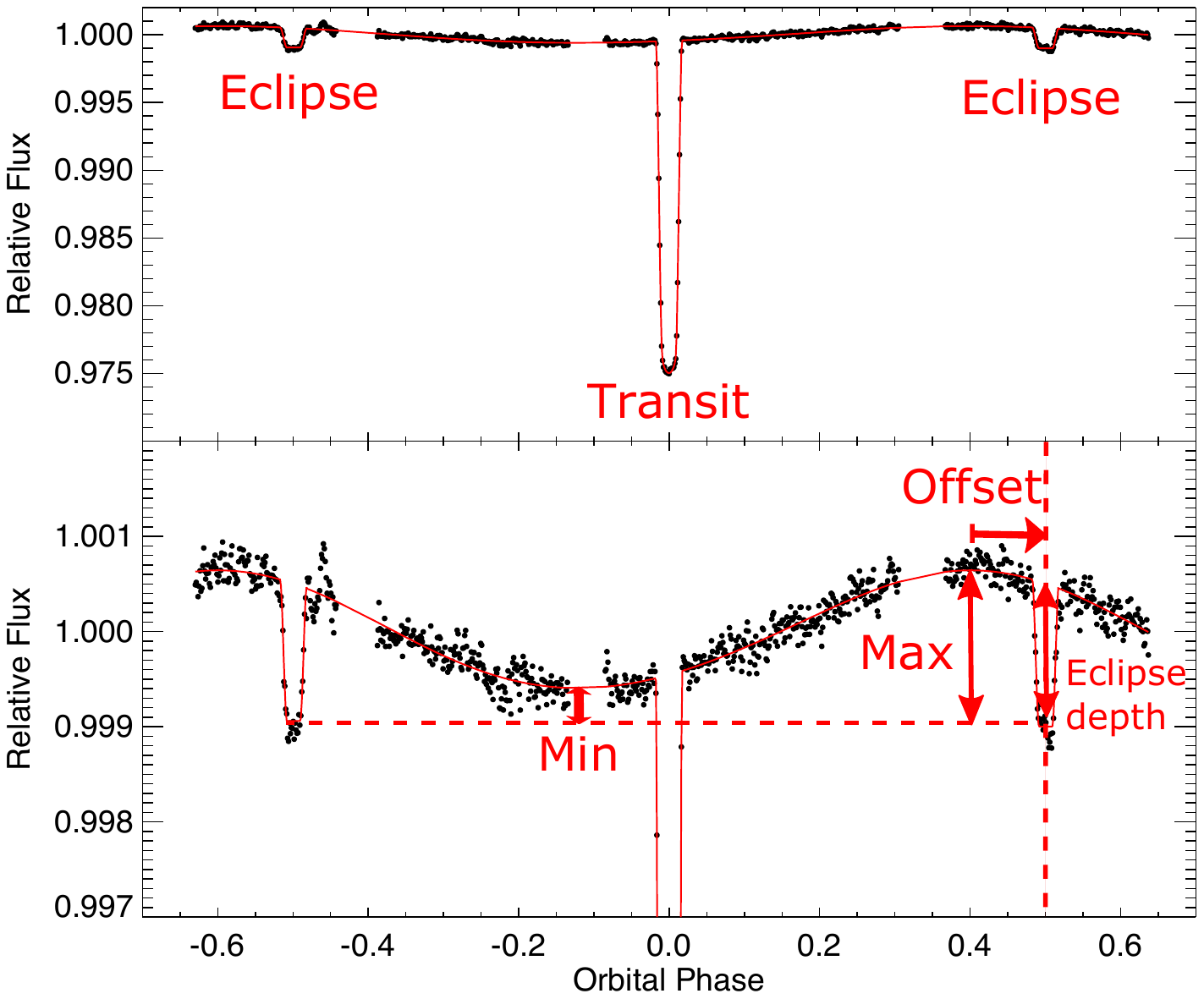}

\caption{Phase curve of HD189733b observed at 3.6$\mu$m with the Spitzer Space Telescope by~\citet{Knutson2012}. The bottom panel is a zoomed-in version of the top panel, with the quantities of interest annotated.}       
\label{fig::HD189PC}
\end{figure}

\subsection{From phase curves to atmospheric properties}

For a steady-state atmosphere, a phase curve provides the longitudinal variation of the hemispherically averaged brightness of the planet. As such, only large-scale atmospheric structures can be inferred (see chapter by Cowan and Fuji). From current phase curve observations, three parameters can usually be retrieved~\citep[e.g.][]{Cowan2008,Demory2013,Knutson2009,Knutson2012}. For transiting planets, these parameters are the secondary eclipse depth, the phase of the maximum of the phase curve compared to the secondary eclipse (called the phase curve offset), and the phase curve relative amplitude,  $A_{\rm F}=(F_{\rm p,Max}-F_{\rm p,Min})/F_{\rm p,Max}$ (see Fig.~\ref{fig::HD189PC}). 

The secondary eclipse depth gives the brightness of the dayside hemisphere (see chapter by Roi Alonso) and serves as a reference to calculate $A_{\rm F}$.

The {phase} offset gives the longitude of the brightest hemisphere of the planet. A planet with a maximum brightness at the substellar point and a symmetrical brightness distribution will have a phase curve peaking during the secondary eclipse. A planet that is brighter \emph{east} of the substellar point will have a phase curve peaking \emph{before} secondary eclipse, leading, by definition, to a  \emph{positive} offset of the phase curve maximum. A planet that is brighter  \emph{west} of the substellar point will have a phase curve peaking \emph{after} secondary eclipse, leading to a \emph{negative} offset. Here we define \emph{east} and \emph{west} with respect to the rotation of the planet. For tidally locked planets, the rotation and the revolution have the same direction and the link between phase curve offset and brightness distribution is independent of the observer's position. 

The phase curve relative amplitude $A_{\rm F}$ provides insight into the brightness contrast between the brightest and the dimmest hemisphere. It goes from 0 (corresponding to no brightness variation) to 1 (when one hemisphere emits zero flux). Different flavors of this parameter can be used. As an example,~\citet{Perez-Becker2013a} defined the day/night relative amplitude ($F_{\rm p,Day}-F_{\rm p,Night})/F_{\rm p,Day}$), whereas~\citet{Komacek2016} used the day/night brightness temperature contrast. The day/night contrast is useful when studying the energy balance of the atmosphere~\citep{Cowan2012,Schwartz2015}, but its estimated value can depend on the measurement of the phase curve offset~\citep{Schwartz2017}. The temperature contrast might seem a more intuitive quantity, however, for phase curves sparsely sampling the spectral space, the conversion from flux to temperatures is model dependent~\citep{Cowan2012}.

For massive planets, gravitational interactions between the planet and the star can lead to non-atmospheric signals in the phase curve such as the ellipsoidal variations and the beaming effect~\citep{Shporer2017}. Although techniques can be used to isolate the atmospheric component of the phase curve, it inevitably leads to higher uncertainties in the derived atmospheric properties~\citep{Shporer2015}.

\subsection{Observational Challenges}

Phase curves are a uniquely challenging phenomenon to observe.  The
timescale and characteristic shape of phase curves make them much more
difficult to observe than occultations.  While the magnitude of a
phase curve signal is comparable to that of the more commonly observed
transits or eclipse, the phase curve timescale $\sim$days is much
longer than the $\sim$hour timescale of occultations.  Furthermore,
phase curves contain only low-frequency components \citep{Cowan2008},
while the high-frequency component is much more prominent in transits
and eclipses.  Occultation observations can be safely decorrelated
even against high-order polynomials or complicated and/or periodic
signals \citep[e.g.,][]{Haynes2015}, which is much more challenging
for phase curves.

In contrast, a single phase curve observation can be compromised by
just a single monotonic trend. For example, it took five years to
recognize that $\sim$40\% of the originally published Spitzer/IRAC
8\,$\mu$m phase curve of HD~189733b was compromised by systematics
\citep{Knutson2007,Knutson2012}.  Similarly, the Spitzer/MIPS
24\,$\mu$m phase curve of HD~209458b was irretrievably compromised by
an instrumental drift lasting tens of hours \citep{Crossfield2012b}.
In such analyses, as with the case of the apparent ellipsoidal
variation seen around WASP-12b at 4.5\,$\mu$m
\citep{Cowan2012,Stevenson2014a}, the question is often: how
does one distinguish signal from systematic?

The most straightforward solution is probably also the most effective:
repeat the experiment, and observe multiple phase curves at the same
wavelengths.  This is the approach adopted in the first HST/WFC3
1.1--1.7\,$\mu$m {\em spectroscopic} phase curve, of WASP-43b
\citep{Stevenson2014b}. By observing three phase curves, a
simultaneous analysis allowed the phase curve to be fit while
separating out a long quadratic systematic in addition to the standard
HST effects. While multiple observations with the same instrument do
not guarantee repeatability \citep{Stevenson2017}, this approach at
least ensures that questionable results are identified.  Beyond
observing multiple phase curves, a more fundamental requirement is to
at least observe a full phase curve -- i.e., begin and end with a
secondary eclipse -- whenever possible.

Implicit in the above comments is the assumption of a
continuously observed phase curve. The first phase curve observations
employed a more parsimonious, ``snapshot'' approach using as few as
five discrete-observing epochs \citep{Harrington2006}. This analysis
was followed by a combined, snapshot plus continuous, program that
measured a mid-infrared phase curve and confirmed the long-term stability of
the MIPS 24$\,\mu$m detector in most cases
\citep{Crossfield2010}. Later, with the benefit of over a decade of
Spitzer analyses, \cite{Krick2016} demonstrated the ability of
Spitzer/IRAC to also obtain snapshot-mode phase curves. Caution may be
advised since the phase curve amplitudes derived from the IRAC
snapshot and from continuous observations \citep{Wong2015} differ by
$\sim$5$\sigma$. However, the snapshot technique is not obviously
flawed -- even multiple continuous phase curves of the same target
sometimes disagree by up to $\sim$4$\sigma$ \citep{Stevenson2017}.
Regardless, the snapshot technique is unlikely to dominate in the JWST
era owing to the large overheads associated with even modest telescope
slews.

\subsection{Wavelength dependence}

Observing exoplanet phase curves at different wavelengths is essential to obtain a complete view of their atmosphere. At optical wavelengths, where the brightness of the planet is often dominated by reflected light, the phase curve provides information on the longitudinal variation of the planet's albedo. At infrared wavelengths, where the brightness of the planet is often dominated by thermal emission, the phase curve provides information on the longitudinal variation of the planet's temperature and chemical composition. 

The spectral variation of {thermal phase curves} is shaped by molecular features. Inside a molecular absorption band, a phase curve probes low pressures (high altitudes) while probing deeper outside a molecular absorption band \citep{Showman2009,Kataria2015}. As a consequence, {multiwavelength phase curves} probe 2D (longitude, depth) thermal and chemical structure of the atmosphere~\citep{Knutson2009,Stevenson2014b,Stevenson2017}. In general, there is no reason to believe that phase curves obtained in different bandpasses will be similar. When looking for trends between the amplitude or offset of phase curves and planetary parameters, it is therefore important to compare observations taken in the same bandpasses. 

\begin{figure}
\includegraphics[width=0.5\linewidth]{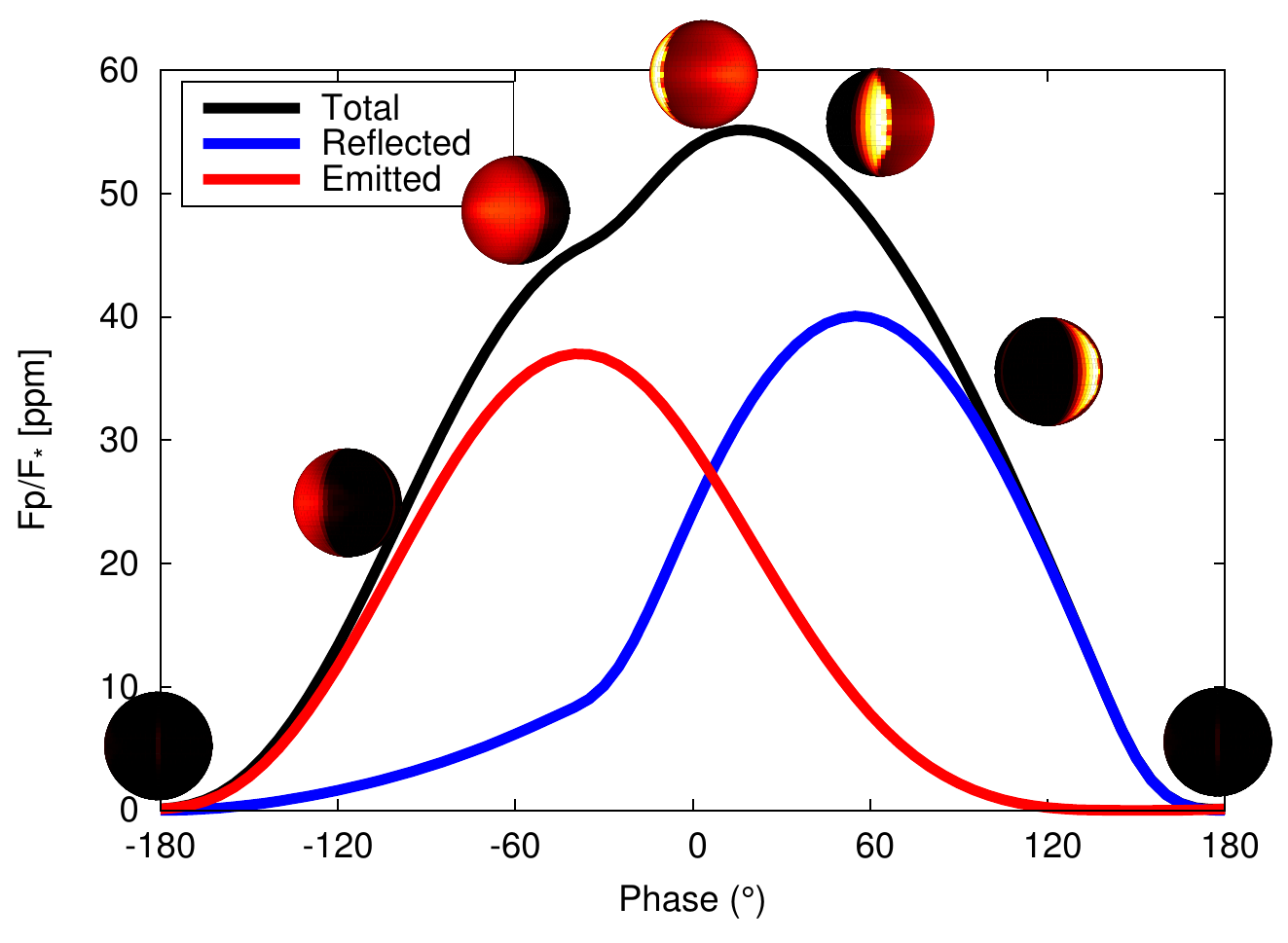}
\includegraphics[width=0.5\linewidth]{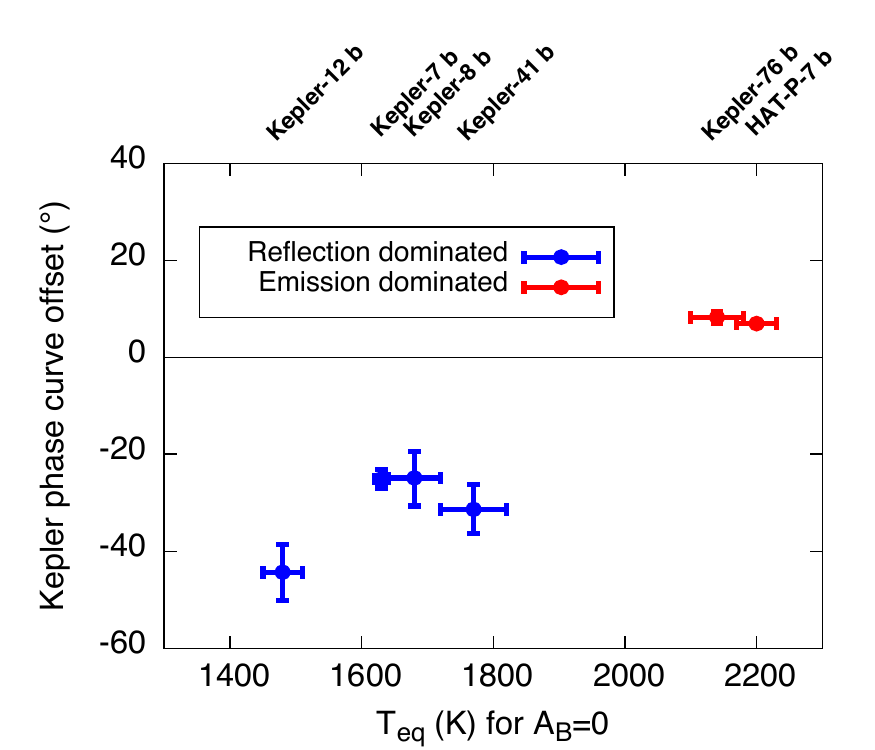}

\caption{\emph{Left} : Phase curve in the Kepler bandpass calculated from a hot Jupiter model with $T_{\rm eq}=1900K$ including silicate clouds. The phase curve (black)  is a combination of the reflected phase curve (blue) and the thermal emission phase curve (red). The circles are brightness maps of the Earth-facing hemisphere. The bright strip on the east of the substellar point is reflected light due to clouds whereas the rest is thermal emission.  \emph{Right} : Observed offsets in the Kepler phase curve sample. Planets with a small equilibrium temperature, dominated by reflected light, have a negative offset, whereas planets with a high equilibrium temperature, dominated by thermal emission, have a positive offset. Figure from~\citet{Parmentier2016}.}

\label{fig::KeplerPC}       
\end{figure}

In some bandpasses, both reflected light and thermal emission can contribute significantly. Fig.~\ref{fig::KeplerPC} shows the theoretical reflected and thermal phase curves of a hot Jupiter in the Kepler bandpass (400-800nm). The reflected phase curve is dominated by the presence of bright clouds west of the substellar point, whereas the thermal emission phase curve is dominated by the presence of a temperature maximum east of the substellar point. For this specific model ($T_{\rm eq}=1900\rm K$) and bandpass, thermal emission and reflected light compensate, and no offset is measured. For cooler planets, reflected light should dominate over thermal emission, whereas for hotter planets it should be the opposite, leading to the correlation between phase curve offset and equilibrium temperature for Kepler phase curves seen in the right panel of Fig.~\ref{fig::KeplerPC}.

\subsection{Summary of observations to date  }

\begin{table}
\caption{Current exoplanets phase curve observations}
\label{tab:1}       
\begin{tabular}{p{1.8cm}p{1cm}p{1cm}p{1cm}p{1.8cm}p{1.6cm}p{2.9cm}}
\hline\noalign{\smallskip}
Planet & $T_{\rm eq}^e (K)$& P (days)& $\lambda$ ($\mu m$) & $\rm A_{\rm F}$ & Offset ($^{\circ}$) & Reference  \\
\noalign{\smallskip}\svhline\noalign{\smallskip}
55 CnC e &1958&0.74& $4.5$&$0.76\pm0.27$&$41\pm12$&~\citet{Demory2016}\\[2pt]
\hline
GJ436b$^g$ &770&2.64& $8$ &$0.40\pm0.13$&$4\pm26$&~\citet{Stevenson2012}\\[2pt]
\hline 
HAT-P-7b &2226&2.2& $0.6$ & $1.02\pm0.04$$^c$&$7.9\pm0.3$$^a$&~\citet{Esteves2015}\\
                &&& $3.6$  & $0.8\pm0.15$&$6.8\pm7.5$&~\citet{Wong2016}\\
                &&& $4.5$ & $0.6\pm0.11$&$4.1\pm7.5$ &~\citet{Wong2016}\\[2pt]
 \hline
HD149026b&1645&2.87 & $8$& $0.70\pm0.34$&n/a&~\citet{Knutson2009}\\[2pt]
\hline
HD189733b &1198&2.218& $3.6$     & $0.845\pm0.061 $  & $35.8\pm4$     & ~\citet{Knutson2012}\\[2pt]
                    &&&    $4.5$  & $0.55\pm0.06 $  & $20.1\pm5.5$  & ~\citet{Knutson2012}\\[2pt]
                    &&&      $8$  & $>0.096$        & $23.7\pm2.7$  & ~\citet{Knutson2012}\\[2pt]
                    &&&    $24$  & $0.26\pm0.08$   & $37.2\pm8.1$  & ~\citet{Knutson2012}\\[2pt]
\hline                    
HD209458b &1445&3.52& $4.5$  &$0.71\pm0.05$ & $40.9\pm6$& ~\citet{Zellem2014}\\[2pt]
\hline
$\upsilon$ And b$^f$&1233&4.62&$24$&n/a&$84.5\pm6.3$ &~\citet{Crossfield2010}\\[2pt]
\hline
WASP-12b &2990&1.09& $3.6$ & $ 0.90\pm0.19$$^{b}$ &$53\pm7$$^d$&~\citet{Cowan2012}\\[2pt]
      		  &&& $4.5$ &$ 0.95\pm0.12$$^{b}$&$16\pm4$&~\citet{Cowan2012}\\[2pt]
\hline
WASP-14b  &1875&2.24& $3.6$      & $>0.91$ & $9.56\pm1.4$ &~\citet{Wong2015}\\[2pt]
                    &&&$4.5$       &$0.70\pm0.06$ & $6.75\pm1.4$&~\citet{Wong2015}\\[2pt]
\hline
WASP-18b &2396&0.94&$1.5$&$>0.93$&$4.5\pm$0.5&~\citet{Arcangeli2017}\\[2pt]
		  &&& $3.6$ & $0.97\pm0.09$ &$0.3\pm1$& ~\citet{Maxted2013a}$^b$\\[2pt]
		  &&& $4.5$ & $ 0.96\pm0.06$ &$-3.6\pm2.3$  &~\citet{Maxted2013a}$^b$\\[2pt]
\hline 
WASP-19b&2064&0.79& $3.6 $ & $ 0.96\pm0.11$ &$10.45\pm4$ &~\citet{Wong2016}\\[2pt]
		  &&&  $4.5$ & $0.81\pm0.12$ & $12.9\pm3.6$&~\citet{Wong2016}\\[2pt]
\hline
WASP-43b  &1441&0.81&$1.5$      &$1.005\pm0.013$ & $12.3\pm1.0$ & ~\citet{Stevenson2014b}\\[2pt]
		   &&& $3.6$     & $1\pm0.03$        & $12.2\pm1.7$  &~\citet{Stevenson2017}\\[2pt]
                   &&& $4.5$      &  $1\pm0.03 $       &$21\pm1.8$     &~\citet{Stevenson2017}\\[2pt]
\hline
WASP-103b  &2508&0.92&$1.5$      &$0.91\pm0.01$ & $-0.3\pm0.1$ & ~\citet{Kreidberg2018}\\[2pt]
		   &&& $3.6$     & $0.91\pm0.13$        & $2\pm0.7$  &~\citet{Kreidberg2018}\\[2pt]
                   &&& $4.5$      &  $0.83\pm0.05$       &$1 \pm0.4$     &~\citet{Kreidberg2018}\\[2pt]
                 
\hline
Kepler-7b &1630&4.88& $0.6$& $1.12\pm0.38$$^c$&$-25\pm1.9$&~\citet{Esteves2015}\\[2pt]
\hline
Kepler-8b &1680&3.52& $0.6$& $0.74\pm0.45$$^c$&$-24.8\pm5.6$&~\citet{Esteves2015}\\[2pt]
\hline
Kepler-10b &2120 &	0.83 &$0.6$&$1.12\pm0.27$&$9\pm6$&~\citet{Hu2015}\\[2pt]
\hline
Kepler-12b&1480&4.43& $0.6$&$ 0.89\pm0.26$$^c$&$-44.3\pm5.8$&~\citet{Esteves2015}\\[2pt]
\hline
Kepler-41b&1770&1.86& $0.6$&$ 1.30\pm0.48$$^c$&$-31\pm5$    &~\citet{Esteves2015}\\[2pt]
\hline
Kepler-76b&2140&1.54& $0.6$ &$0.80\pm0.05$$^c$&$8.3\pm1.2$&~\citet{Esteves2015}\\[2pt]
\noalign{\smallskip}\hline\noalign{\smallskip}
\end{tabular}
For optical phase curves, only planets for which an offset is detected have been considered. The phase curves of highly eccentric planets are discussed in a later section.\\
Only planets on circular orbits have been taken into account here, a discussion of eccentric planets can be found in the devoted section.\\
Instruments used are Spitzer for the 3.6, 4.5, 8 and 24$\mu m$ observations, HST/WFC3-IR for the $1.5\mu m$ observations, Kepler for the 0.6$\mu m$ observations\\
$^a$ The offset varies from orbit to orbit~\citep{Armstrong2016}
$^b$ $A_{\rm F}$ values were taken from ~\citet{Stevenson2017}
$^c$ Uncertainty for phase curve maximum observed in the Kepler bandpass was assumed to be similar to the uncertainty of the secondary eclipse.
$^d$ Using a different noise model the same authors get an offset of $0\pm29$.\\
$^e$ We use $T_{\rm eq}^4=\frac{T_{\rm star}^4R_{\rm star}^2}{4a^2}$, $a$ : semi-major axis, $T_{\rm star}$, $R_{\rm star}$ : stellar effective temperature and radius.
$^f$ This is a non transiting planet with a measured inclination of $24\pm4^{\circ}$~\citep{Piskorz2017}.
$^g$ This planet has a non-zero but small eccentricity (e=0.1371)~\citep[see][]{Stevenson2010}.
\label{tab::Obs}
\end{table}

\begin{figure}
\includegraphics[width=\linewidth]{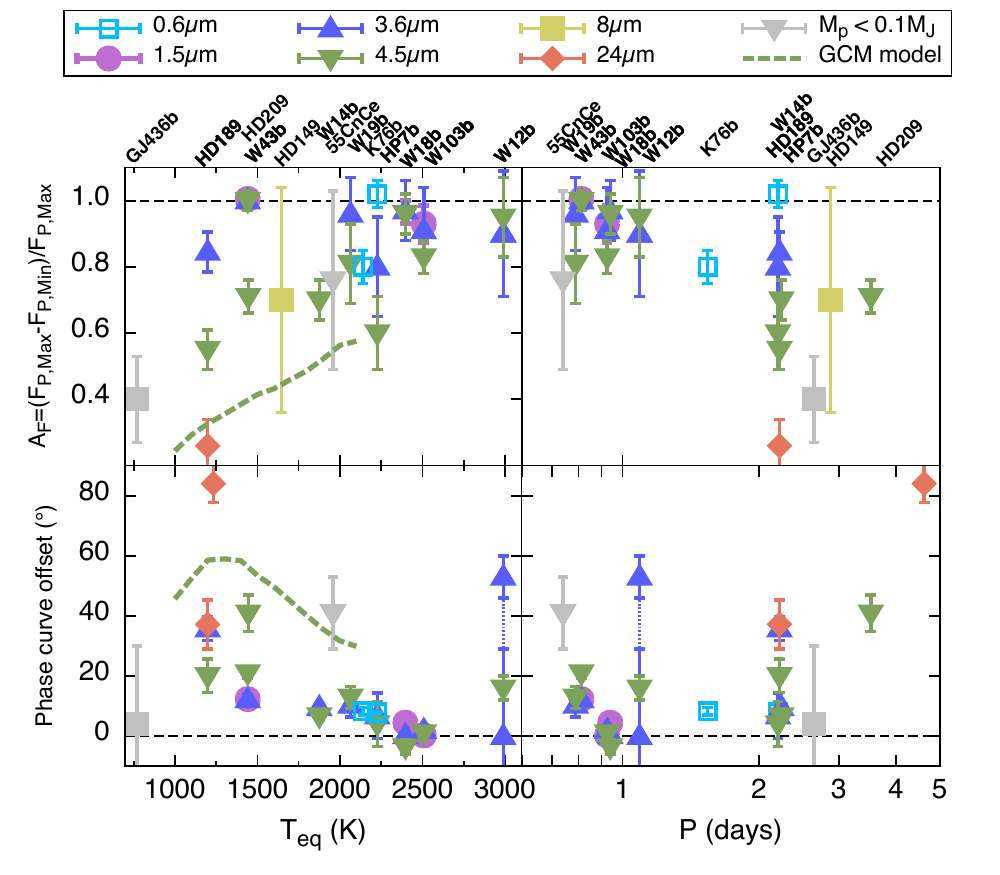}

\caption{Amplitude (top) and offset (bottom) of exoplanet thermal phase curves as a function of equilibrium temperature (left) and orbital period (right). Different symbols and colors represent different wavelengths. Values predicted at $4.5\mu m$ by the Jupiter radius, solar composition, chemical equilibrium, cloudless and drag-free global circulation model of~\citet{Parmentier2016} are overplotted. The model systematically underpredicts the amplitude and overpredicts the offset of the phase curve. For the $0.6\mu m$ case, only planets with a phase curve dominated by thermal emission are considered (i.e. planets with $T_{\rm eq}>2000\rm K$). For WASP-12b, the two possible offsets derived from the same phase curve are linked with a dashed curve~\citep{Cowan2012}. Colored points are hot Jupiters, gray points are super-Earths and mini-Neptunes. The data is from Table 1.}       
\label{fig::Obs}
\end{figure}

Table~\ref{tab::Obs} lists the current phase curve observations of transiting exoplanets, while Fig.~\ref{fig::Obs} shows the observations for the thermal phase curves of hot Jupiters. In general, hot Jupiters' thermal phase curves have a large amplitude (between 0.5 and 1) and have a positive offset, meaning that the brightest hemisphere is eastward of the substellar point. 

As seen in Fig.~\ref{fig::Obs}, there is no clear trend between the amplitude of thermal phase curves and planet equilibrium temperature. Earlier claims by~\citet[][]{Perez-Becker2013a} and~\citet{Komacek2016} were based on a smaller number of observations and by interpreting together observations taken in different bandpasses. In today's more complete dataset, no trend is seen, neither by looking at all bandpasses together nor by looking at them separately. 

A tentative trend, first proposed by~\citet{Stevenson2017}, is seen in the amplitude vs. planet rotation period plot: planets with a faster rotation rate might have a larger phase curve amplitude. Observations of planets with an orbital period between 1 and 2 days and, especially, larger than 3 days are currently being taken to confirm the presence of a correlation (Spitzer program 13038).

The most striking feature of Fig.~\ref{fig::Obs}  might be in the phase curve offset vs. equilibrium temperature plot, with a lack of large  offset for planets with equilibrium temperatures  $\gtrsim 1700$~K. WASP-12b seems an outlier, but it's large offset might be the result of uncorrected instrument systematics~\citep{Cowan2012}. The super-Earth 55~Cnc~e does not fit in this trend defined by hot Jupiter observations. Its phase curve appears similar to cooler hot Jupiters, such as HD209458b, which has a large offset and a large phase curve amplitude at $4.5\mu m$, pointing toward unique atmospheric or ground properties. 

\section{The physics of exoplanet phase curves}
\subsection{Thermal structure}
\subsubsection{Atmospheric regimes}
In response to the large day-to-night irradiation gradient, tidally locked planets develop a strong {atmospheric circulation} that advects energy from the dayside to the nightside and reduces the day/night temperature contrast compared to the local radiative equilibrium case. However, energy losses by radiation or by the damping of the winds can prevent a fully efficient redistribution of energy. This balance between energy transport and energy losses can be understood in terms of timescales~\citep{Showman2002,Komacek2016}. The {radiative timescale} is the characteristic time it takes for a parcel of gas to lose its energy by radiation. The {wave timescale} is the time it takes for a gravity wave to travel horizontally over one planetary radius. The {drag timescale} is the time it takes for the waves or the winds to lose a significant part of their kinetic energy. 

Depending on the balance between these three timescales, the atmospheric circulation on tidally locked planets is expected to follow at least two different regimes (see also chapter by N. Lewis). When the radiative timescale and the drag timescales are similar or longer than the wave timescale, the presence of an eastward, equatorial jet is expected. The resulting temperature map is shifted eastward compared to a local radiative equilibrium case, and the hottest spot of the atmosphere lays eastward of the substellar point. When the radiative timescale or the drag timescale is smaller than the wave timescale, the atmospheric circulation is characterized by a day-to-night flow with a temperature map symmetric around the substellar point and a somewhat larger hemispheric temperature contrast~\citep[e.g.][]{Showman2011,Tsai2014}.

\subsubsection{Phenomenological models vs. global circulation models}

The {energy balance} of a planet can be determined by measuring the thermal emission from two opposite hemispheres. When considering the day- and nightside hemispheres, one can understand the energy balance in terms of how much of the incoming energy is reflected to space (i.e., the Bond albedo) and how much of the absorbed stellar light is transferred to the nightside (i.e., the redistribution factor). In order to have a complete energy balance, the full dayside and nightside spectra should be used. However, current observations only cover a few bandpasses, and the effective temperature of each hemisphere must be estimated by extrapolating the measurements with assumptions that are only exact if each hemisphere emits like a blackbody. These approximations should work for the planet's dayside~\citep{Cowan2011} but might fail when estimating the nightside temperatures~\citep{Schwartz2015} where the vertical temperature gradient is larger. The latest results from \citet{Schwartz2017} show a large dispersion in both redistribution efficiency and geometric albedo between planets, with no apparent trend between equilibrium temperature and day/night temperature contrast, echoing the lack of trends seen in the top left panel of Fig.~\ref{fig::Obs}.~\citet{Schwartz2017} also conclude that many hot Jupiters have Bond albedos larger than $0.3$, in apparent contradiction with their measured low geometric albedos in the Kepler bandpass~\citep{Heng2013}. This apparent discrepancy could be the result of an asymmetric scattering function~\citep[e.g.,][]{Dyudina2005}, a lower geometric albedo in the Kepler bandpass than outside of it~\citep{Crossfield2015}, or an intrinsic bias in the method. 

Linking the phase curve amplitude and the phase offset is a more complex task that requires a model of the longitudinal distribution of the temperature and the opacities. One (and two) -dimensional models of the longitudinal (and latitudinal) variation of the temperature have been calculated by taking into account the competing effects of longitudinal advection of energy and radiative losses~\citep{Cowan2011a,Hu2015,Zhang2017}. They lead to a temperature map that is determined by one parameter: the ratio of the advective and the radiative timescale. Consequently, they always predict a correlation between the phase curve offset and the phase curve amplitude. These models, however, lack vertical transport, which is proven to be an important factor setting the day/night temperature contrast~\citep{Komacek2016}.  The phase curve offset and the phase curve amplitude might therefore be set by different mechanisms, which could be the reason why they are not correlated in the  observations~\citep{Crossfield2015} nor in the more complex, three-dimensional models of tidally locked exoplanets~\citep{Komacek2017, Parmentier2016}.

Whereas these phenomenological models are useful to retrieve parameters such as the radiative timescale from phase curve observations, they can't say much about the mechanisms setting these parameters. Global circulation models, solving for the hydrodynamics, the radiative transfer, and/or the magnetic effects in three dimensions, have been used to quantitatively link the observed day/night contrast and phase curve offset to planetary parameters. As seen in Fig.~\ref{fig::Obs}, cloudless, dragless, solar composition models systematically underpredict the amplitude and overpredict the offset of thermal phase curves~\citep{Showman2009,Kataria2015}. Other mechanisms, such as metallicity~\citep{Kataria2015}, cloud composition~\citep{Oreshenko2016,Parmentier2016}, rotation period~\citep{Showman2009, Showman2011}, disequilibrium chemistry~\citep{Cooper2006}, or magnetic field strength~\citep{Rogers2017} must play an important role. 

\subsubsection{Radiative timescale}

The {radiative timescale} is expected to vary linearly with pressure and with the inverse of the cube of the temperature~\citep{Iro2005,Showman2008}. Since the pressures probed by atmospheric sensing span several orders of magnitude, the pressure dependence of the radiative timescale has a major consequence on the observations. Phase curves obtained at wavelengths probing deep in the atmosphere are expected to probe layers with a large radiative timescale and thus a small amplitude and a large offset, whereas phase curves obtained at wavelengths probing shallower layers (e.g., inside molecular bands) should have a larger amplitude and smaller offset. This is illustrated in the cloudless case of Fig.~\ref{fig::LambdaDependence}. 

The temperature dependence of the radiative timescale is also expected to impact the observations. With all things being equal, hotter planets are expected to cool more efficiently, leading to a larger {day/night temperature contrast} and thus a larger phase curve amplitude~\citep{Perez-Becker2013a,Komacek2016}. As seen in Fig.~\ref{fig::Obs}, the lack of a clear trend in the amplitude vs. equilibrium temperature plot indicates that other mechanisms must contribute to the shape of exoplanets' phase curves~\citep{Komacek2017}. 

\subsubsection{Drag, composition and mean molecular weight}

The thermal structure of the planet can be affected when either the radiative timescale, the drag timescale, or the wave timescales are changed. 

\begin{figure}
\includegraphics[scale=1]{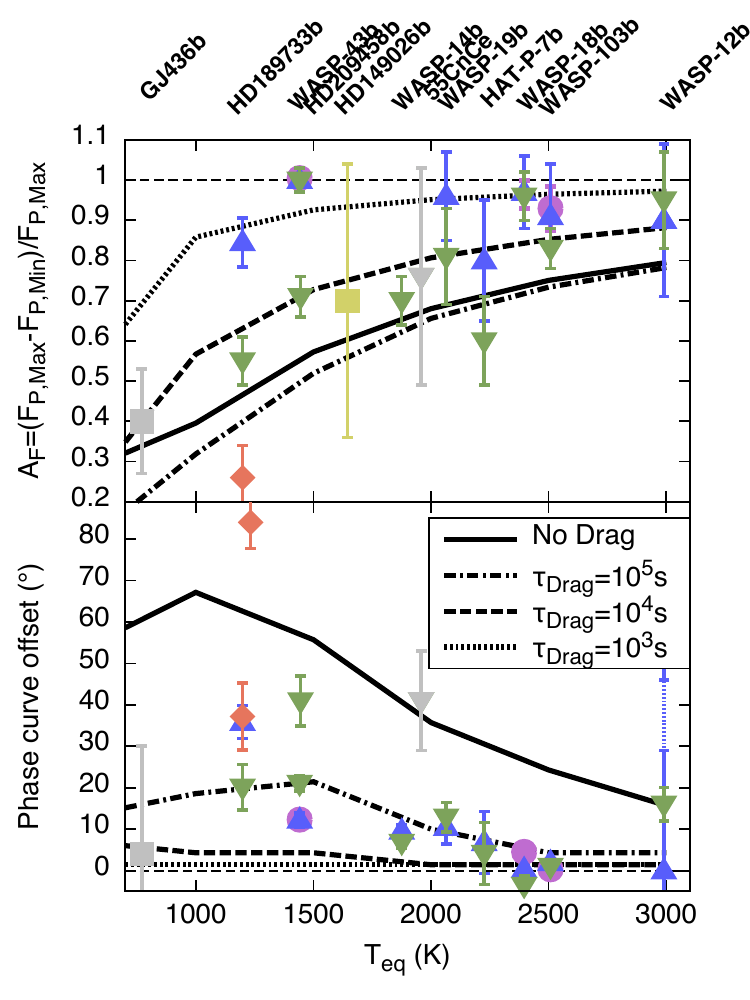}
\caption{Observed thermal phase curve amplitude (upper panel) and offset of the maximum (bottom panel) compared with semi-gray global circulation models from~\citep{Komacek2017} using different strength of drag. The color and shape of the points are the same as in Fig.~\ref{fig::Obs}.}
\label{fig::Drag}       
\end{figure}

{Atmospheric drag} is often used and parametrized as Rayleigh drag in models to represent different types of physical mechanisms such as shocks~\citep{Perna2012,Heng2012a}, hydrodynamic instabilities~\citep{Fromang2016} or Lorentz forces~\citep{Perna2010,Batygin2013}. Its main effect is to weaken the super-rotating jet, leading to a more pronounced substellar to anti-stellar flow and a weaker equatorial eastward jet. As seen in Fig.~\ref{fig::Drag}, small Rayleigh drag ($\tau_{\rm drag}\approx 10^5$~s) has a small effect on the phase curve amplitude but a large effect on the phase offset. By shifting the circulation pattern from a jet-dominated to a day-to-night flow, a small drag does not change much the heat transport to the nightside but reduces the longitudinal asymmetry of the temperature and therefore the phase curve offset. A larger drag (e.g., $\tau_{\rm drag}\approx10^3-10^4$~s) can increase the phase curve amplitude significantly, but lead to an extremely small phase curve offset. It is important to keep in mind that the physical mechanism responsible for the drag might not be homogeneous in the atmosphere or might not even act as a Rayleigh drag at all. For example, Lorenz forces should have a fundamentally different effect on the flow than Rayleigh drag, leading to a potential inversion of the equatorial jet speed~\citep{Rogers2014a,Rogers2017}, whereas hydrodynamic instabilities and shocks should drag the flow where large velocity or temperature gradients are present~\citep{Fromang2016,Heng2012a}.

Atmospheric composition can also play an important role in shaping the phase curve of exoplanets. By increasing the abundance of metals in the atmosphere, {metallicity} increases the opacities, leading to two consequences. First the atmospheric layers probed are shifted to smaller pressures, where the radiative timescale is smaller leading to a large phase curve amplitude and a smaller phase curve offset. Second the enhanced opacities lead to a warmer atmosphere and thus to a smaller radiative timescale. Overall, increasing the atmospheric metallicity should increase the amplitude and decrease the offset of the phase curve. For WASP-43b, global circulation models with a metallicity of five times solar provide a better, but not satisfactory, match to the observations compared to models assuming a solar metallicity~\citep{Kataria2015}.

In planets with equilibrium temperatures larger than $\approx 1900\rm K$, titanium and vanadium oxide are expected to be present in the dayside atmosphere and create a strong thermal inversion~\citep{Hubeny2003,Fortney2008,Parmentier2015,Parmentier2016}. The detection of such an inverted temperature profile has recently been claimed through the presence of molecular emission features in the dayside of both WASP-33b~\citep{Haynes2015} and WASP-121b~\citep{Evans2017}. Thermal inversions are expected to disappear in the planet's nightside where these molecular features are expected to become absorption features.

In small planets, where the atmospheric composition can be very diverse, the mean molecular weight of the atmosphere is expected to impact both the radiative timescale and the wave timescale.~\citet{Zhang2017} show that the day/night temperature contrast is expected to increase, and the eastward shift of the hot spot is expected to decrease, when the mean molecular weight is increased. In real planets, atmospheres with different mean molecular weight will have a very different composition, leading to variations of several orders of magnitude in the opacities and potentially a comparable or stronger effect than the effect of the mean molecular weight alone.  

\subsection{Opacity structure}

\subsubsection{Chemistry}
In chemical equilibrium, temperature inhomogeneities often pair up with chemical composition inhomogeneities. For a given atomic composition, equilibrium favors different molecules at different pressures and temperatures. A day/night temperature contrast on a tidally locked planet could result in a day/night chemical gradient. Such a large-scale change in the chemistry should trigger a change in the opacities and affect the offset and the amplitude of the phase curve. Horizontal and vertical advection are expected to further complicate this picture. When the chemical reaction timescale is long compared to the horizontal and the vertical mixing timescale, chemical reactions cannot happen fast enough during the transport of gas from one side to the other side of the planet and quenching happens. The atmospheric circulation can then drive the chemical abundances out of their local equilibrium state and erase any chemical gradient expected from the local equilibrium~\citep{Cooper2006,Visscher2011}. The exact chemical composition of the atmosphere is determined by a combination of the chemical equilibrium abundances of the deep atmosphere (vertical quenching) and of the hot dayside (horizontal quenching)~\citep{Agundez2014}. Chemical {quenching} is expected to affect the opacities in very specific bands, leading to a peculiar signature in the day/night contrast vs. wavelength relationship~\citep{Steinrueck2017}.

\begin{figure}
\includegraphics[width=\linewidth]{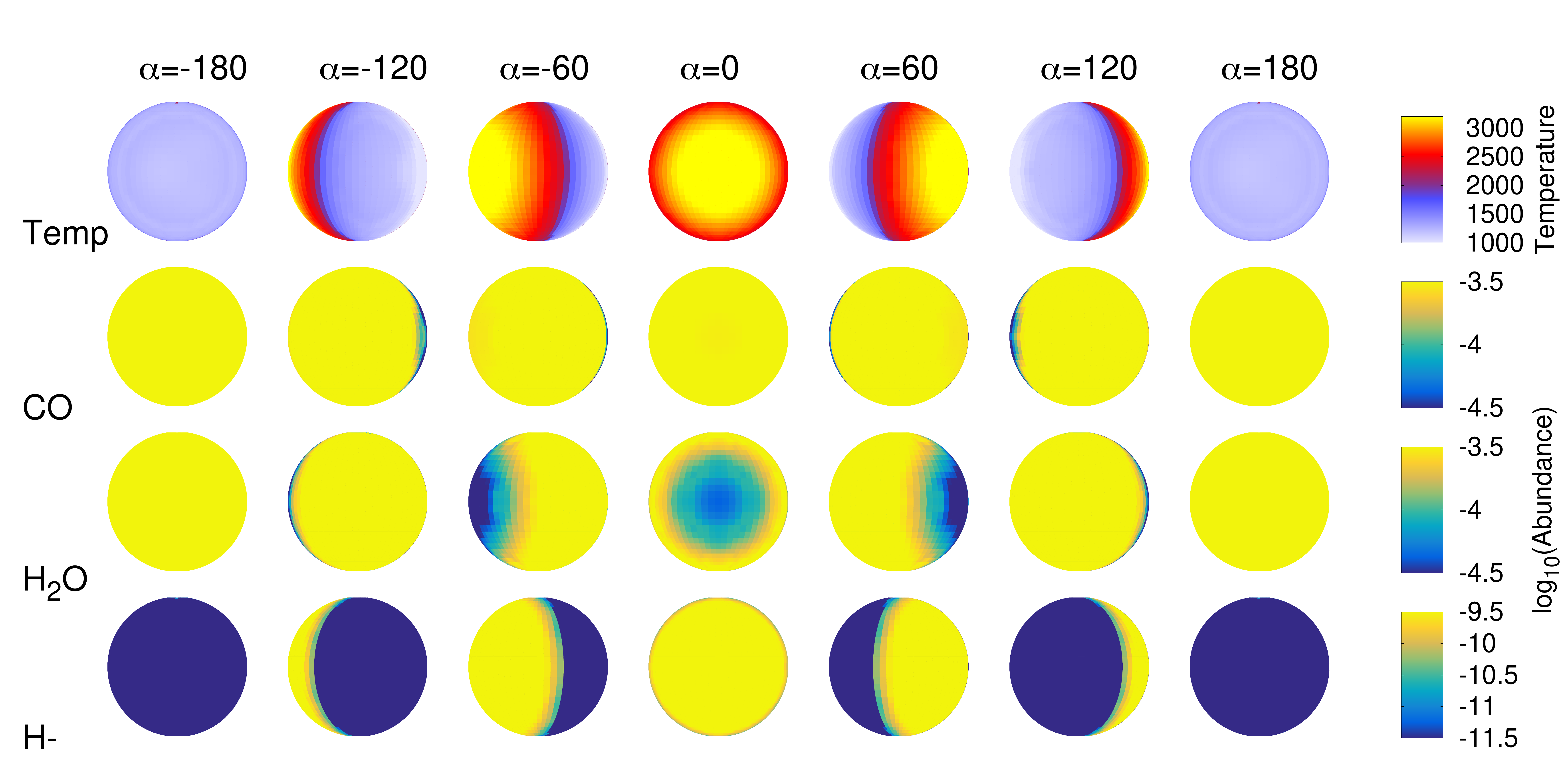}
\caption{Temperature (top) and abundances of water, CO and H$\rm-$ at the $1.4\mu m$ photosphere of WASP-103b as predicted from global circulation models of~\citet{Parmentier2018}. The different columns are for different orbital phases, $\alpha=0$ being the secondary eclipse. The water abundance drops by two orders of magnitude in the dayside due to thermal dissociation~\citep{Kreidberg2018}. The abundances are calculated assuming local chemical equilibrium~\citep{Visscher2006}.}
\label{fig::W103b}       
\end{figure}

In very hot planets ($T_{\rm eq}\gtrsim 2500\rm K$) molecules such as water should be thermally dissociated at the dayside photosphere but not in the nightside, whereas more strongly bond molecules, such as CO, should be present at all longitudes (see Fig.~\ref{fig::W103b}). When molecules dissociate, continuum opacities from hydrogen ions are expected to become dominant, leading to a lack of spectral features and a blackbody-like thermal emission in the dayside~\citep{Kreidberg2018, Arcangeli2017}.

\subsubsection{Clouds}

The large temperature variations in the atmosphere of tidally locked planets in close-in orbits are responsible for the prevalence of large longitudinal inhomogeneities in their {cloud} coverage as inferred from the observations~\citep{Demory2013,Shporer2015}. In each planet, some species can condense in the cold nightside of the planet but must be in gaseous form in the hot dayside~\citep{Parmentier2013}. Since the temperature map of most hot Jupiters is not symmetric, but shifted eastward, the western part of the dayside is usually cold and clouds can also form there. The reflected phase curve can then be dominated by the bright reflective part of the dayside atmosphere where cloud are present, leading to a phase curve peaking after secondary eclipse, the opposite sign than for thermal phase curve (see~\citet{Hu2015,GarciaMunoz2015,Webber2015} and Fig~\ref{fig::KeplerPC}). The longitudinal distribution of a given type of cloud in the dayside is determined to first order by the thermal structure of the planet and the cloud-specific condensation temperature~\citep{Lee2016}. If the thermal structure is known, either by observing a thermal phase curve or by modeling the temperature distribution a priori, the cloud map derived from the reflected light curve can be used to constrain the cloud chemical composition~\citep{Oreshenko2016,Parmentier2016}. 

\begin{figure}
\includegraphics[scale=0.6]{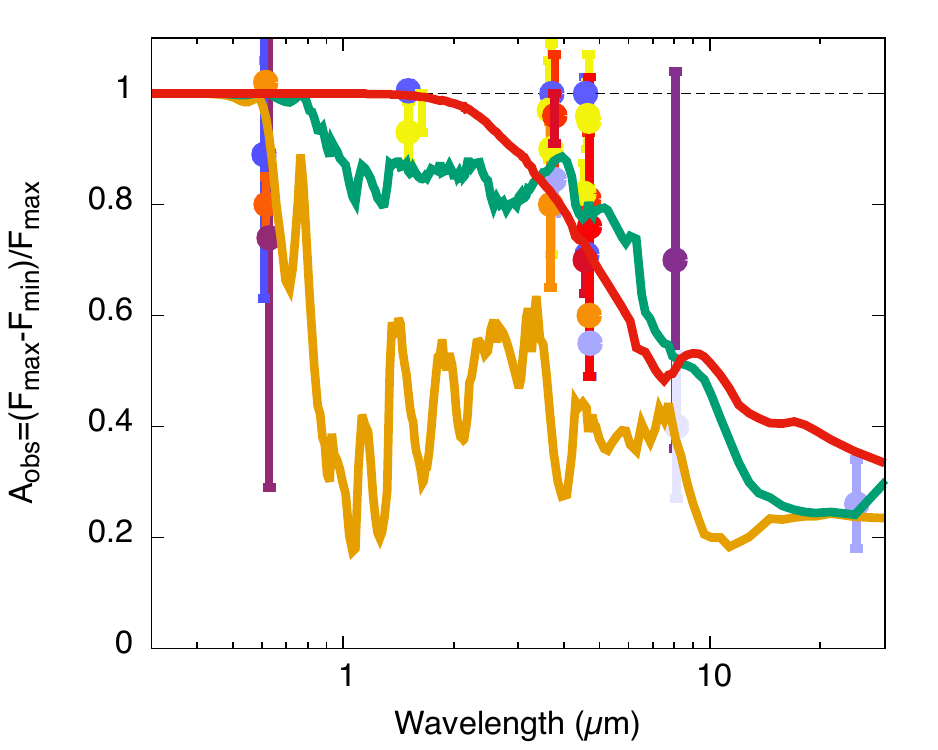}
\includegraphics[scale=0.6]{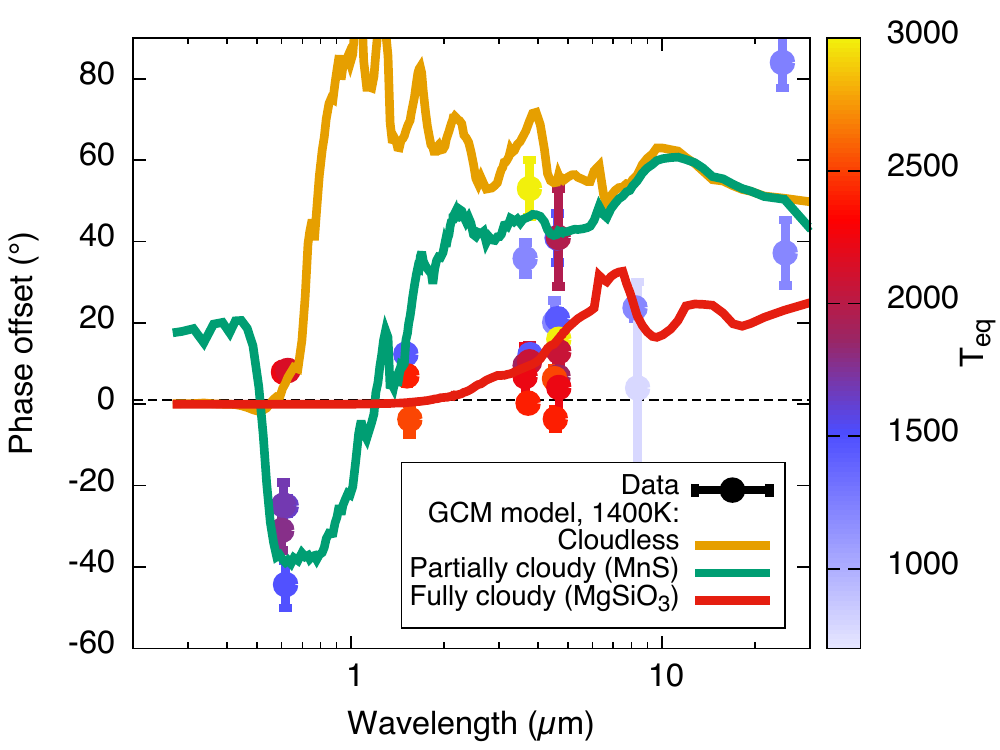}

\caption{Relative amplitude (left) and phase offset (right) of currently known exoplanet phase curves as a function of wavelength. The color of the points represents the equilibrium temperature of the planet. Predictions from three different global circulation models of a hot Jupiter with $T_{\rm eq}=1400K$ have been overplotted~\citep{Parmentier2017}: an example of a cloudless model (orange), a model with a partial cloud coverage (green, assuming the presence of MnS clouds) and an example of a fully cloudy model (red, assuming the presence of $\rm MgSiO_3$).}
\label{fig::LambdaDependence}       
\end{figure}

Clouds should also affect the thermal emission of the planet. The presence of {nightside clouds} produces a large opacity gradient between the day and the night that can suppress the thermal emission from the cloudy regions by raising the photosphere to low pressures. As a consequence, clouds are expected to increase the phase curve amplitude and decrease the phase curve offset of hot Jupiters (see Fig.~\ref{fig::LambdaDependence}), even if they are present only on the planet's nightside. When clouds are present, the brightest hemisphere is not necessarily the hottest one, and the shift of the maximum of thermal phase curves does not track the shift of the hottest point of the planet anymore~\citep{Parmentier2017}. Similarly, the day/night flux contrast cannot be easily converted to a day/night horizontal temperature contrast, a redistribution parameter or an advective timescale. Any inference of the thermal structure and atmospheric composition of an atmosphere based on the phase offset and the day/night contrast of a thermal phase curve can be highly biased by the presence of clouds. An optical phase curve, probing the planet albedo as a function of longitude seems a necessary complement to any thermal phase curve.

\subsection{Time variability}

Atmospheres can be time variable, leading to a time dependence of the phase curve amplitude and offset between orbits. Given the large number of observations needed to observe one phase curve and thus the time needed to constrain differences between phase curves, atmospheric variability has not yet been widely characterized. The most notable exception is the planet HAT-P-7b, a hot Jupiter that has been observed continuously by the Kepler spacecraft over more than 700 orbital periods. As shown by~\citet{Armstrong2016}, the shift of the maximum of the phase curve may vary from positive to negative values over a timescale of 50 orbits. Two scenarios have been proposed to explain this behavior. In the first scenario, a global change in the dayside temperature of the planet, without necessarily a change in the position of the hot spot, would lead to a variation in the cloud coverage, changing the relative ratio of thermal and reflected contribution to the light curve \citep{Armstrong2016}. In the second scenario, proposed by~\citet{Rogers2017}, the entire phase curve of HAT-P-7b is dominated by thermal emission. The coupling between the ionized atmosphere and the magnetic field leads to an oscillation in the strength and direction of the equatorial jet and the center of the hottest hemisphere itself oscillates from east to west of the substellar point. Observations of variability at infrared wavelength should be able to disentangle between the two scenarios. In the first case, no variation is expected at infrared wavelengths, where reflected light is never dominant. In the second scenario, the change of the hottest hemisphere longitude should affect similarly the Kepler phase curve and the infrared phase curve. 

\subsection{Eccentric planets}
\label{par::eccentric}
Highly {eccentric planets} are laboratories to understand the response of a planet atmosphere to a time-dependent forcing. Quantities that are degenerate in the phase curves of tidally locked planets, such as the radiative timescale and the advective timescale, can potentially be measured separately in the eccentric case~\citep{Lewis2013,deWit2016}. When passing through periastron, the planet is inhomogeneously and transiently heated. Then, depending of the rotation period of the planet~\citep{Hut1981}, the heated hemisphere is expected to face Earth periodically after periastron, leading observable oscillations (or "ringing") in the phase curve~\citep{Langton2008,Kataria2013}. While the period of these oscillations are related to the rotation period of the planet, their amplitude is linked to the radiative timescale of the atmosphere~\citep{Cowan2011a}. So far, the thermal phase curves of two highly eccentric planets have been published, while the phase curve of a third one (XO-3b) has been observed but is not yet published. No oscillations were found in the thermal phase curves of HAT-P-2b and HD80606b, which was interpreted by the presence of a short ($\approx 2-6\rm h$) radiative timescale at the photosphere~\citep{Lewis2013,deWit2016}. Such a short radiative timescale is reproduced by global circulation models of HAT-P-2b~\citep{Langton2008,Lewis2013} but not by models of HD80606b, where the cooler temperatures predict a longer timescale. Any opacity source, such as clouds, could raise the photosphere of HD80606b to lower pressures and account for this short radiative timescale~\citep{Lewis2017}. Planned Hubble Space Telescope observations should measure the albedo of HD80606b and its variation around periastron.

\subsection{Non-transiting planets}

Phase curve observations can be used to characterize the atmospheres of non-transiting planets on short orbits. The number of non-transiting planets is far greater than the number of transiting ones, so {non-transiting planets} can potentially provide a large sample of bright planets to study~\citep{Millholland2017}. Unfortunately, the phase curve relative amplitude cannot be determined directly for non-transiting, even when the inclination of the system is known, as both the radius of the planet and the stellar absolute flux cannot be measured directly~\citep{Crossfield2010}. The phase curve offset, however, can be directly determined from the observation, given a good enough knowledge of the planet ephemeris. So far,  $\upsilon$~And~b is the only non-transiting planet with a known inclination~\citep{Piskorz2017} that has a measured phase curve offset~\citep{Crossfield2010}. Measuring a phase curve offset for planets over a wide range of inclinations could be used to infer the latitudinal thermal structure of exoplanets' atmospheres. 

\section{Future prospects with JWST}
The James Webb Space Telescope ({JWST}) will provide valuable information on giant tidally locked planets through phase curve observations. By observing both optical infrared wavelengths, NIRISS should be able, in one spectroscopic phase curve observation, to provide both the cloud map and the temperature map of hot Jupiters and show the change of sign of the phase curve offset with wavelength expected for a partially cloudy atmosphere (see Fig.~\ref{fig::LambdaDependence}). NIRCAM and NIRSPEC will provide valuable measurements of molecular abundances and their variation with longitude. These near-infrared wavelengths, however, are not best suited to observe the cooler nightside of tidally locked planets where most observations measured a close-to-zero flux and could not detect any molecular features. These cooler nightsides should, however, be brighter at longer wavelengths, where MIRI can observe. 

{CHEOPS}, {TESS}, and {PLATO} will significantly increase the number of optical phase curves of hot Jupiters, giving important constraint on the cloud distribution on exoplanets~\citep[see][for a complete review]{Shporer2017}. Moreover, these missions will find numerous planets, particularly in the Neptune size range, that current and future observatories (HST, JWST) will be able to characterize. As such, numerous phase curves of Neptune and sub-Neptune size planets could be obtained during the next decade, increasing drastically the current sample of two (55 CnC e and GJ436b, see Figure~\ref{fig::Obs}).

{FINESSE} and {ARIEL}, currently into the selection process, would provide hundreds of exoplanets spectroscopic phase curves covering a wide range of planet mass, radius, irradiation and rotation period. Such a survey mission is needed to understand the diversity of climates on extrasolar worlds. 

JWST will probably be used to search for the phase curves of temperate rocky planets orbiting M dwarfs~\citep{Kreidberg2016}, such as the Trappist-1 planets, Proxima centauri b or LHS 1140b. Theoretically, the phase curve amplitude of these systems can be used to determine whether the planet possesses a fluid that can efficiently transport energy (i.e., an atmosphere or an ocean), since the amplitude of the phase curve of a bare rock would be larger than any other cases. In practice, these observations will be extremely challenging as they require an exquisite stability of the telescope and the star over a period of days to tens of days. Future and more precise observatories, such as the {Origins Space Telescope} in the infrared or {LUVOIR} in the optical, will be more likely to obtain phase curve measurements of temperate planets. 

Interpreting further the shape of these temperate rocky planet phase curves will be complex, as the number of unconstrained parameters that can influence the phase curve of a rocky planet is larger than the ones influencing a hot Jupiter, for which we already face many challenges. Among them are the large range of possible atmospheric composition~\citep{Leconte2015}, the possibility of asynchronous rotation~\citep{Leconte2015a}, the range of possible atmospheric mass~\citep{Koll2015}, the thermal inertia of the ground~\citep{Selsis2013}, the peculiar behavior of clouds~\citep{Yang2013}, the effects of non-dilute atmospheres~\citep{Pierrehumbert2016} or the intertwined signal of different planets (including undetected ones) on the same system~\citep{Kane2013}. 

\section{Conclusion}

Exoplanet phase curves are fundamental probes of the three-dimensional structure of exoplanets in close-in orbits. The current sample shows that every exoplanet studied so far is unique, making the search for trends between phase curve shape and planetary parameters harder. State-of-the-art models are unable to quantitatively explain current phase curve observations. Whether clouds, drag, atmospheric variability, chemistry, or instrumental systematics are behind the unexpectedly large amplitude and small offsets of hot Jupiters, thermal phase curves should be determined in the near future through the wavelength dependence of the phase curve shape as will be observed by coming spacecrafts. Numerous observations of cooler and smaller planets are expected in the next decade and will probably challenge our current understanding of planetary atmospheres.

\section{Cross-References}
\begin{itemize}
\item{Observing Exoplanets with the Spitzer Space Telescope}
\item{Characterization of Exoplanets: Secondary Eclipses}
\item{Mapping Exoplanets}
\item{Exoplanet Atmosphere Observations from Transmission Spectroscopy and Other Planet-Star Combined Light Observational Techniques}
\item{Atmospheric Circulation for Exoplanet Atmospheres}
\item{Radiative Transfer for Exoplanet Atmospheres}
\end{itemize}

\begin{acknowledgement}
We thanks Kevin Stevenson for useful feedback on the manuscript. V.P. acknowledges support from the Sagan Postdoctoral Fellowship through the NASA Exoplanet Science Institute. 
\end{acknowledgement}
\bibliographystyle{spbasicHBexo}  
\bibliography{Parmentier_Phase_ArXiv.bbl} 

\begin{thebibliography}{90}
\providecommand{\natexlab}[1]{#1}
\providecommand{\url}[1]{{#1}}
\providecommand{\urlprefix}{URL }
\expandafter\ifx\csname urlstyle\endcsname\relax
  \providecommand{\doi}[1]{DOI~\discretionary{}{}{}#1}\else
  \providecommand{\doi}{DOI~\discretionary{}{}{}\begingroup
  \urlstyle{rm}\Url}\fi
\providecommand{\eprint}[2][]{\url{#2}}

\bibitem[{{Ag\'undez} et~al.(2014){Ag\'undez}, {Parmentier}, {Venot},
  {Hersant}, and {Selsis}}]{Agundez2014}
{Ag\'undez} M, {Parmentier} V, {Venot} O, {Hersant} F {Selsis} F (2014) Pseudo
  2d chemical model of hot-jupiter atmospheres: application to hd 209458b and
  hd 189733b. A\&A 564:A73,
  \urlprefix\url{http://dx.doi.org/10.1051/0004-6361/201322895}

\bibitem[{{Arcangeli} et~al.(2017){Arcangeli}, {Desert}, {Parmentier}, {Line},
  {Bean}, and {Fortney}}]{Arcangeli2017}
{Arcangeli} J, {Desert} JM, {Parmentier} V et~al. (2017) Climate of a massive
  gas giant: Phase-curve of wasp-18b with hst. in prep

\bibitem[{Armstrong et~al.(2016)Armstrong, de~Mooij, Barstow, Osborn, Blake,
  and Saniee}]{Armstrong2016}
Armstrong DJ, de~Mooij E, Barstow J et~al. (2016) Variability in the atmosphere
  of the hot giant planet hat-p-7 b. Nature Astronomy 1:0004 EP --,
  \urlprefix\url{http://dx.doi.org/10.1038/s41550-016-0004}

\bibitem[{{Batygin} et~al.(2013){Batygin}, {Stanley}, and
  {Stevenson}}]{Batygin2013}
{Batygin} K, {Stanley} S {Stevenson} DJ (2013) {Magnetically Controlled
  Circulation on Hot Extrasolar Planets}. \apj 776:53

\bibitem[{{Cooper} and {Showman}(2006)}]{Cooper2006}
{Cooper} CS {Showman} AP (2006) {Dynamics and Disequilibrium Carbon Chemistry
  in Hot Jupiter Atmospheres, with Application to HD 209458b}. \apj
  649:1048--1063

\bibitem[{{Cowan} and {Agol}(2011{\natexlab{a}})}]{Cowan2011}
{Cowan} NB {Agol} (2011{\natexlab{a}}) {The Statistics of Albedo and Heat
  Recirculation on Hot Exoplanets}. \apj 729:54

\bibitem[{{Cowan} and {Agol}(2008)}]{Cowan2008}
{Cowan} NB {Agol} E (2008) {Inverting Phase Functions to Map Exoplanets}. \apjl
  678:L129--L132

\bibitem[{{Cowan} and {Agol}(2011{\natexlab{b}})}]{Cowan2011a}
{Cowan} NB {Agol} E (2011{\natexlab{b}}) {A Model for Thermal Phase Variations
  of Circular and Eccentric Exoplanets}. \apj 726:82

\bibitem[{{Cowan} et~al.(2012){Cowan}, {Machalek}, {Croll}, {Shekhtman},
  {Burrows}, {Deming}, {Greene}, and {Hora}}]{Cowan2012}
{Cowan} NB, {Machalek} P, {Croll} B et~al. (2012) {Thermal Phase Variations of
  WASP-12b: Defying Predictions}. \apj 747:82

\bibitem[{{Crossfield}(2015)}]{Crossfield2015}
{Crossfield} IJM (2015) {Observations of Exoplanet Atmospheres}. \pasp
  127:941--960

\bibitem[{{Crossfield} et~al.(2010){Crossfield}, {Hansen}, {Harrington}, {Cho},
  {Deming}, {Menou}, and {Seager}}]{Crossfield2010}
{Crossfield} IJM, {Hansen} BMS, {Harrington} J et~al. (2010) {A New 24 {$\mu$}m
  Phase Curve for {$\upsilon$} Andromedae b}. \apj 723:1436--1446

\bibitem[{{Crossfield} et~al.(2012){Crossfield}, {Knutson}, {Fortney},
  {Showman}, {Cowan}, and {Deming}}]{Crossfield2012b}
{Crossfield} IJM, {Knutson} H, {Fortney} J et~al. (2012) {Spitzer/MIPS 24
  {$\mu$}m Observations of HD 209458b: Three Eclipses, Two and a Half Transits,
  and a Phase Curve Corrupted by Instrumental Sensitivity Variations}. \apj
  752:81

\bibitem[{{de Wit} et~al.(2016){de Wit}, {Lewis}, {Langton}, {Laughlin},
  {Deming}, {Batygin}, and {Fortney}}]{deWit2016}
{de Wit} J, {Lewis} NK, {Langton} J et~al. (2016) {Direct Measure of Radiative
  and Dynamical Properties of an Exoplanet Atmosphere}. \apjl 820:L33

\bibitem[{{Demory} et~al.(2013){Demory}, {de Wit}, {Lewis}, {Fortney}, {Zsom},
  {Seager}, {Knutson}, {Heng}, {Madhusudhan}, {Gillon}, {Barclay}, {Desert},
  {Parmentier}, and {Cowan}}]{Demory2013}
{Demory} BO, {de Wit} J, {Lewis} N et~al. (2013) {Inference of Inhomogeneous
  Clouds in an Exoplanet Atmosphere}. \apjl 776:L25

\bibitem[{{Demory} et~al.(2016){Demory}, {Gillon}, {de Wit}, {Madhusudhan},
  {Bolmont}, {Heng}, {Kataria}, {Lewis}, {Hu}, {Krick}, {Stamenkovi{\'c}},
  {Benneke}, {Kane}, and {Queloz}}]{Demory2016}
{Demory} BO, {Gillon} M, {de Wit} J et~al. (2016) {A map of the large day-night
  temperature gradient of a super-Earth exoplanet}. \nat 532:207--209

\bibitem[{{Dyudina} et~al.(2005){Dyudina}, {Sackett}, {Bayliss}, {Seager},
  {Porco}, {Throop}, and {Dones}}]{Dyudina2005}
{Dyudina} UA, {Sackett} PD, {Bayliss} DDR et~al. (2005) {Phase Light Curves for
  Extrasolar Jupiters and Saturns}. \apj 618:973--986

\bibitem[{{Esteves} et~al.(2015){Esteves}, {De Mooij}, and
  {Jayawardhana}}]{Esteves2015}
{Esteves} LJ, {De Mooij} EJW {Jayawardhana} R (2015) {Changing Phases of Alien
  Worlds: Probing Atmospheres of Kepler Planets with High-precision
  Photometry}. \apj 804:150

\bibitem[{{Evans} et~al.(2017){Evans}, {Sing}, {Kataria}, {Goyal}, {Nikolov},
  {Wakeford}, {Deming}, {Marley}, {Amundsen}, {Ballester}, {Barstow},
  {Ben-Jaffel}, {Bourrier}, {Buchhave}, {Cohen}, {Ehrenreich}, {Garc{\'{\i}}a
  Mu{\~n}oz}, {Henry}, {Knutson}, {Lavvas}, {Etangs}, {Lewis},
  {L{\'o}pez-Morales}, {Mandell}, {Sanz-Forcada}, {Tremblin}, and
  {Lupu}}]{Evans2017}
{Evans} TM, {Sing} DK, {Kataria} T et~al. (2017) {An ultrahot gas-giant
  exoplanet with a stratosphere}. \nat 548:58--61

\bibitem[{{Feng} et~al.(2016){Feng}, {Line}, {Fortney}, {Stevenson}, {Bean},
  {Kreidberg}, and {Parmentier}}]{Feng2016}
{Feng} YK, {Line} MR, {Fortney} JJ et~al. (2016) {The Impact of Non-uniform
  Thermal Structure on the Interpretation of Exoplanet Emission Spectra}. \apj
  829:52

\bibitem[{{Fortney} et~al.(2008){Fortney}, {Lodders}, {Marley}, and
  {Freedman}}]{Fortney2008}
{Fortney} JJ, {Lodders} K, {Marley} MS {Freedman} RS (2008) {A Unified Theory
  for the Atmospheres of the Hot and Very Hot Jupiters: Two Classes of
  Irradiated Atmospheres}. \apj 678:1419--1435

\bibitem[{{Fromang} et~al.(2016){Fromang}, {Leconte}, and {Heng}}]{Fromang2016}
{Fromang} S, {Leconte} J {Heng} K (2016) {Shear-driven instabilities and shocks
  in the atmospheres of hot Jupiters}. \aap 591:A144

\bibitem[{{Garcia Munoz} and {Isaak}(2015)}]{GarciaMunoz2015}
{Garcia Munoz} A {Isaak} KG (2015) {Probing exoplanet clouds with optical phase
  curves}. ArXiv e-prints

\bibitem[{{Guillot} et~al.(1996){Guillot}, {Burrows}, {Hubbard}, {Lunine}, and
  {Saumon}}]{Guillot1996}
{Guillot} T, {Burrows} A, {Hubbard} WB, {Lunine} JI {Saumon} D (1996) {Giant
  Planets at Small Orbital Distances}. \apjl 459:L35

\bibitem[{{Harrington} et~al.(2006){Harrington}, {Hansen}, {Luszcz}, {Seager},
  {Deming}, {Menou}, {Cho}, and {Richardson}}]{Harrington2006}
{Harrington} J, {Hansen} BM, {Luszcz} SH et~al. (2006) {The Phase-Dependent
  Infrared Brightness of the Extrasolar Planet {$\upsilon$} Andromedae b}. Science
  314:623--626

\bibitem[{{Haynes} et~al.(2015){Haynes}, {Mandell}, {Madhusudhan}, {Deming},
  and {Knutson}}]{Haynes2015}
{Haynes} K, {Mandell} AM, {Madhusudhan} N, {Deming} D {Knutson} H (2015)
  {Spectroscopic Evidence for a Temperature Inversion in the Dayside Atmosphere
  of Hot Jupiter WASP-33b}. \apj 806:146

\bibitem[{{Heng}(2012)}]{Heng2012a}
{Heng} K (2012) {On the Existence of Shocks in Irradiated Exoplanetary
  Atmospheres}. \apjl 761:L1

\bibitem[{{Heng} and {Demory}(2013)}]{Heng2013}
{Heng} K {Demory} BO (2013) {Understanding Trends Associated with Clouds in
  Irradiated Exoplanets}. \apj 777:100

\bibitem[{{Hu} et~al.(2015){Hu}, {Demory}, {Seager}, {Lewis}, and
  {Showman}}]{Hu2015}
{Hu} R, {Demory} BO, {Seager} S, {Lewis} N {Showman} AP (2015) {A
  Semi-analytical Model of Visible-wavelength Phase Curves of Exoplanets and
  Applications to Kepler- 7 b and Kepler- 10 b}. \apj 802:51

\bibitem[{{Hubeny} et~al.(2003){Hubeny}, {Burrows}, and
  {Sudarsky}}]{Hubeny2003}
{Hubeny} I, {Burrows} A {Sudarsky} D (2003) {A Possible Bifurcation in
  Atmospheres of Strongly Irradiated Stars and Planets}. \apj 594:1011--1018

\bibitem[{{Hut}(1981)}]{Hut1981}
{Hut} P (1981) {Tidal evolution in close binary systems}. \aap 99:126--140

\bibitem[{{Iro} et~al.(2005){Iro}, {B{\'e}zard}, and {Guillot}}]{Iro2005}
{Iro} N, {B{\'e}zard} B {Guillot} T (2005) {A time-dependent radiative model of
  HD 209458b}. \aap 436:719--727

\bibitem[{{Kane} and {Gelino}(2013)}]{Kane2013}
{Kane} SR {Gelino} DM (2013) {Decoupling Phase Variations in Multi-planet
  Systems}. \apj 762:129

\bibitem[{{Kataria} et~al.(2013){Kataria}, {Showman}, {Lewis}, {Fortney},
  {Marley}, and {Freedman}}]{Kataria2013}
{Kataria} T, {Showman} AP, {Lewis} NK et~al. (2013) {Three-dimensional
  Atmospheric Circulation of Hot Jupiters on Highly Eccentric Orbits}. \apj
  767:76

\bibitem[{{Kataria} et~al.(2015){Kataria}, {Showman}, {Fortney}, {Stevenson},
  {Line}, {Kreidberg}, {Bean}, and {D{\'e}sert}}]{Kataria2015}
{Kataria} T, {Showman} AP, {Fortney} JJ et~al. (2015) {The Atmospheric
  Circulation of the Hot Jupiter WASP-43b: Comparing Three-dimensional Models
  to Spectrophotometric Data}. \apj 801:86

\bibitem[{{Knutson} et~al.(2007){Knutson}, {Charbonneau}, {Allen}, {Fortney},
  {Agol}, {Cowan}, {Showman}, {Cooper}, and {Megeath}}]{Knutson2007}
{Knutson} HA, {Charbonneau} D, {Allen} LE et~al. (2007) {A map of the day-night
  contrast of the extrasolar planet HD 189733b}. \nat 447:183--186

\bibitem[{{Knutson} et~al.(2009){Knutson}, {Charbonneau}, {Cowan}, {Fortney},
  {Showman}, {Agol}, {Henry}, {Everett}, and {Allen}}]{Knutson2009}
{Knutson} HA, {Charbonneau} D, {Cowan} NB et~al. (2009) {Multiwavelength
  Constraints on the Day-Night Circulation Patterns of HD 189733b}. \apj
  690:822--836

\bibitem[{{Knutson} et~al.(2012){Knutson}, {Lewis}, {Fortney}, {Burrows},
  {Showman}, {Cowan}, {Agol}, {Aigrain}, {Charbonneau}, {Deming}, {D{\'e}sert},
  {Henry}, {Langton}, and {Laughlin}}]{Knutson2012}
{Knutson} HA, {Lewis} N, {Fortney} JJ et~al. (2012) {3.6 and 4.5 {$\mu$}m Phase
  Curves and Evidence for Non-equilibrium Chemistry in the Atmosphere of
  Extrasolar Planet HD 189733b}. \apj 754:22

\bibitem[{{Koll} and {Abbot}(2015)}]{Koll2015}
{Koll} DDB {Abbot} DS (2015) {Deciphering Thermal Phase Curves of Dry, Tidally
  Locked Terrestrial Planets}. \apj 802:21

\bibitem[{{Komacek} and {Showman}(2016)}]{Komacek2016}
{Komacek} TD {Showman} AP (2016) {Atmospheric Circulation of Hot Jupiters:
  Dayside/Nightside Temperature Differences}. \apj 821:16

\bibitem[{{Komacek} et~al.(2017){Komacek}, {Showman}, and {Tan}}]{Komacek2017}
{Komacek} TD, {Showman} AP {Tan} X (2017) {Atmospheric Circulation of Hot
  Jupiters: Dayside/Nightside Temperature Differences. II. Comparison
  with Observations}. \apj 835:198

\bibitem[{{Kreidberg} and {Loeb}(2016)}]{Kreidberg2016}
{Kreidberg} L {Loeb} A (2016) {Prospects for Characterizing the Atmosphere of
  Proxima Centauri b}. \apjl 832:L12

\bibitem[{{Kreidberg} et~al.(2018){Kreidberg}, {Line}, {Parmentier},
  {Stevenson}, {Louden}, {Bonnefoy}, {Faherty}, {Henry}, {Williamson},
  {Stassun}, {Beatty}, {Bean}, {Fortney}, {Showman}, {D{\'e}sert}, and
  {Arcangeli}}]{Kreidberg2018}
{Kreidberg} L, {Line} MR, {Parmentier} V et~al. (2018) {Global Climate and
  Atmospheric Composition of the Ultra-hot Jupiter WASP-103b from HST and
  Spitzer Phase Curve Observations}. \aj 156:17

\bibitem[{{Krick} et~al.(2016){Krick}, {Ingalls}, {Carey}, {von Braun}, {Kane},
  {Ciardi}, {Plavchan}, {Wong}, and {Lowrance}}]{Krick2016}
{Krick} JE, {Ingalls} J, {Carey} S et~al. (2016) {Spitzer IRAC Sparsely Sampled
  Phase Curve of the Exoplanet Wasp-14B}. \apj 824:27

\bibitem[{{Langton} and {Laughlin}(2008)}]{Langton2008}
{Langton} J {Laughlin} G (2008) {Hydrodynamic Simulations of Unevenly
  Irradiated Jovian Planets}. \apj 674:1106-1116

\bibitem[{{Leconte} et~al.(2015{\natexlab{a}}){Leconte}, {Forget}, and
  {Lammer}}]{Leconte2015}
{Leconte} J, {Forget} F {Lammer} H (2015{\natexlab{a}}) {On the (anticipated)
  diversity of terrestrial planet atmospheres}. Experimental Astronomy
  40:449--467

\bibitem[{{Leconte} et~al.(2015{\natexlab{b}}){Leconte}, {Wu}, {Menou}, and
  {Murray}}]{Leconte2015a}
{Leconte} J, {Wu} H, {Menou} K {Murray} N (2015{\natexlab{b}}) {Asynchronous
  rotation of Earth-mass planets in the habitable zone of lower-mass stars}.
  Science 347:632--635

\bibitem[{{Lee} et~al.(2016){Lee}, {Dobbs-Dixon}, {Helling}, {Bognar}, and
  {Woitke}}]{Lee2016}
{Lee} G, {Dobbs-Dixon} I, {Helling} C, {Bognar} K {Woitke} P (2016) {Dynamic
  mineral clouds on HD 189733b I. 3D RHD with kinetic, non-equilibrium cloud
  formation}. ArXiv e-prints

\bibitem[{{Lewis} et~al.(2013){Lewis}, {Knutson}, {Showman}, {Cowan},
  {Laughlin}, {Burrows}, {Deming}, {Crepp}, {Mighell}, {Agol}, {Bakos},
  {Charbonneau}, {D{\'e}sert}, {Fischer}, {Fortney}, {Hartman}, {Hinkley},
  {Howard}, {Johnson}, {Kao}, {Langton}, and {Marcy}}]{Lewis2013}
{Lewis} NK, {Knutson} HA, {Showman} AP et~al. (2013) {Orbital Phase Variations
  of the Eccentric Giant Planet HAT-P-2b}. \apj 766:95

\bibitem[{{Lewis} et~al.(2017){Lewis}, {Parmentier}, {Kataria}, {de Wit},
  {Showman}, {Fortney}, and {Marley}}]{Lewis2017}
{Lewis} NK, {Parmentier} V, {Kataria} T et~al. (2017) {Atmospheric Circulation
  and Cloud Evolution on the Highly Eccentric Extrasolar Planet HD 80606b}.
  ArXiv e-prints: 170600466 \urlprefix\url{https://arxiv.org/abs/1706.00466}

\bibitem[{{Line} and {Parmentier}(2016)}]{Line2016}
{Line} MR {Parmentier} V (2016) {The Influence of Nonuniform Cloud Cover on
  Transit Transmission Spectra}. \apj 820:78

\bibitem[{{Maxted} et~al.(2013){Maxted}, {Anderson}, {Doyle}, {Gillon},
  {Harrington}, {Iro}, {Jehin}, {Lafreni{\`e}re}, {Smalley}, and
  {Southworth}}]{Maxted2013a}
{Maxted} PFL, {Anderson} DR, {Doyle} AP et~al. (2013) {Spitzer 3.6 and 4.5
  {$\mu$}m full-orbit light curves of WASP-18}. \mnras 428:2645--2660

\bibitem[{{Millholland} and {Laughlin}(2017)}]{Millholland2017}
{Millholland} S {Laughlin} G (2017) {Supervised Learning Detection of Sixty
  Non-Transiting Hot Jupiter Candidates}. ArXiv e-prints

\bibitem[{{Oreshenko} et~al.(2016){Oreshenko}, {Heng}, and
  {Demory}}]{Oreshenko2016}
{Oreshenko} M, {Heng} K {Demory} BO (2016) {Optical phase curves as diagnostics
  for aerosol composition in exoplanetary atmospheres}. \mnras 457:3420--3429

\bibitem[{{Parmentier} et~al.(2013){Parmentier}, {Showman}, and
  {Lian}}]{Parmentier2013}
{Parmentier} V, {Showman} AP {Lian} Y (2013) {3D mixing in hot Jupiters
  atmospheres. I. Application to the day/night cold trap in HD 209458b}. \aap
  558:A91

\bibitem[{{Parmentier} et~al.(2015){Parmentier}, {Guillot}, {Fortney}, and
  {Marley}}]{Parmentier2015}
{Parmentier} V, {Guillot} T, {Fortney} JJ {Marley} MS (2015) {A non-grey
  analytical model for irradiated atmospheres. II. Analytical vs. numerical
  solutions}. \aap 574:A35

\bibitem[{Parmentier et~al.(2015)Parmentier, Showman, and
  de~Wit}]{Parmentier2015a}
Parmentier V, Showman AP de~Wit J (2015) Unveiling the atmospheres of giant
  exoplanets with an echo-class mission. Experimental Astronomy 40(2):481--500,
  \urlprefix\url{http://dx.doi.org/10.1007/s10686-014-9395-0}

\bibitem[{{Parmentier} et~al.(2016){Parmentier}, {Fortney}, {Showman},
  {Morley}, and {Marley}}]{Parmentier2016}
{Parmentier} V, {Fortney} JJ, {Showman} AP, {Morley} C {Marley} MS (2016)
  {Transitions in the Cloud Composition of Hot Jupiters}. \apj 828:22

\bibitem[{{Parmentier} et~al.(2017){Parmentier}, {Showman}, {Fortney}, and
  {Marley}}]{Parmentier2017}
{Parmentier} V, {Showman} AP, {Fortney} J {Marley} M (2017) The cloudy shape of
  hot jupiters phase curves. in prep

\bibitem[{{Parmentier} et~al.(2018){Parmentier}, {Line}, {Bean}, {Mansfield},
  {Kreidberg}, {Lupu}, {Visscher}, {D{\'e}sert}, {Fortney}, {Deleuil},
  {Arcangeli}, {Showman}, and {Marley}}]{Parmentier2018}
{Parmentier} V, {Line} MR, {Bean} JL et~al. (2018) {From thermal dissociation
  to condensation in the atmospheres of ultra hot Jupiters: WASP-121b in
  context}. \aap 617:A110

\bibitem[{{Perez-Becker} and {Showman}(2013)}]{Perez-Becker2013a}
{Perez-Becker} D {Showman} AP (2013) {Atmospheric Heat Redistribution on Hot
  Jupiters}. \apj 776:134

\bibitem[{{Perna} et~al.(2010){Perna}, {Menou}, and {Rauscher}}]{Perna2010}
{Perna} R, {Menou} K {Rauscher} E (2010) {Magnetic Drag on Hot Jupiter
  Atmospheric Winds}. \apj 719:1421--1426

\bibitem[{{Perna} et~al.(2012){Perna}, {Heng}, and {Pont}}]{Perna2012}
{Perna} R, {Heng} K {Pont} F (2012) {The Effects of Irradiation on Hot Jovian
  Atmospheres: Heat Redistribution and Energy Dissipation}. \apj 751:59

\bibitem[{{Pierrehumbert} and {Ding}(2016)}]{Pierrehumbert2016}
{Pierrehumbert} RT {Ding} F (2016) {Dynamics of atmospheres with a non-dilute
  condensible component}. Proceedings of the Royal Society of London Series A
  472:20160107

\bibitem[{{Piskorz} et~al.(2017){Piskorz}, {Benneke}, {Crockett}, {Lockwood},
  {Blake}, {Barman}, {Bender}, {Carr}, and {Johnson}}]{Piskorz2017}
{Piskorz} D, {Benneke} B, {Crockett} NR et~al. (2017) {Detection of Water Vapor
  in the Thermal Spectrum of the Non-Transiting Hot Jupiter upsilon Andromedae
  b}. ArXiv e-prints

\bibitem[{{Rogers}(2017)}]{Rogers2017}
{Rogers} TM (2017) {Constraints on the magnetic field strength of HAT-P-7 b and
  other hot giant exoplanets}. Nature Astronomy 1:0131

\bibitem[{{Rogers} and {Komacek}(2014)}]{Rogers2014a}
{Rogers} TM {Komacek} TD (2014) {Magnetic Effects in Hot Jupiter Atmospheres}.
  \apj 794:132

\bibitem[{{Schwartz} and {Cowan}(2015)}]{Schwartz2015}
{Schwartz} JC {Cowan} NB (2015) {Balancing the energy budget of short-period
  giant planets: evidence for reflective clouds and optical absorbers}. \mnras
  449:4192--4203

\bibitem[{{Schwartz} et~al.(2017){Schwartz}, {Kashner}, {Jovmir}, and
  {Cowan}}]{Schwartz2017}
{Schwartz} JC, {Kashner} Z, {Jovmir} D {Cowan} NB (2017) {Phase Offsets and the
  Energy Budgets of Hot Jupiters}. ArXiv e-prints

\bibitem[{{Selsis} et~al.(2013){Selsis}, {Maurin}, {Hersant}, {Leconte},
  {Bolmont}, {Raymond}, and {Delbo'}}]{Selsis2013}
{Selsis} F, {Maurin} AS, {Hersant} F et~al. (2013) {The effect of rotation and
  tidal heating on the thermal lightcurves of super Mercuries}. \aap 555:A51

\bibitem[{{Showman} and {Guillot}(2002)}]{Showman2002}
{Showman} AP {Guillot} T (2002) {Atmospheric circulation and tides of ``51
  Pegasus b-like'' planets}. \aap 385:166--180

\bibitem[{{Showman} and {Polvani}(2011)}]{Showman2011}
{Showman} AP {Polvani} LM (2011) {Equatorial Superrotation on Tidally Locked
  Exoplanets}. \apj 738:71

\bibitem[{{Showman} et~al.(2008){Showman}, {Cooper}, {Fortney}, and
  {Marley}}]{Showman2008}
{Showman} AP, {Cooper} CS, {Fortney} JJ {Marley} MS (2008) {Atmospheric
  Circulation of Hot Jupiters: Three-dimensional Circulation Models of HD
  209458b and HD 189733b with Simplified Forcing}. \apj 682:559--576

\bibitem[{{Showman} et~al.(2009){Showman}, {Fortney}, {Lian}, {Marley},
  {Freedman}, {Knutson}, and {Charbonneau}}]{Showman2009}
{Showman} AP, {Fortney} JJ, {Lian} Y et~al. (2009) {Atmospheric Circulation of
  Hot Jupiters: Coupled Radiative-Dynamical General Circulation Model
  Simulations of HD 189733b and HD 209458b}. \apj 699:564--584

\bibitem[{{Shporer}(2017)}]{Shporer2017}
{Shporer} A (2017) {The Astrophysics of Visible-light Orbital Phase Curves in
  the Space Age}. \pasp 129(7):072,001

\bibitem[{{Shporer} and {Hu}(2015)}]{Shporer2015}
{Shporer} A {Hu} R (2015) {Studying Atmosphere-dominated Hot Jupiter Kepler
  Phase Curves: Evidence that Inhomogeneous Atmospheric Reflection Is Common}.
  \aj 150:112

\bibitem[{{Steinrueck} et~al.(2017){Steinrueck}, {Parmentier}, {Showman},
  {Fortney}, and {Lupu}}]{Steinrueck2017}
{Steinrueck} M, {Parmentier} V, {Showman} AP, {Fortney} J {Lupu} R (2017) How
  disequilibrium chemistery affects the phase curve of hot jupiters. in prep

\bibitem[{{Stevenson} et~al.(2010){Stevenson}, {Harrington}, {Nymeyer},
  {Madhusudhan}, {Seager}, {Bowman}, {Hardy}, {Deming}, {Rauscher}, and
  {Lust}}]{Stevenson2010}
{Stevenson} KB, {Harrington} J, {Nymeyer} S et~al. (2010) {Possible
  thermochemical disequilibrium in the atmosphere of the exoplanet GJ 436b}.
  \nat 464:1161--1164

\bibitem[{{Stevenson} et~al.(2012){Stevenson}, {Harrington}, {Lust}, {Lewis},
  {Montagnier}, {Moses}, {Visscher}, {Blecic}, {Hardy}, {Cubillos}, and
  {Campo}}]{Stevenson2012}
{Stevenson} KB, {Harrington} J, {Lust} NB et~al. (2012) {Two nearby
  Sub-Earth-sized Exoplanet Candidates in the GJ 436 System}. \apj 755:9

\bibitem[{{Stevenson} et~al.(2014{\natexlab{a}}){Stevenson}, {Bean},
  {Madhusudhan}, and {Harrington}}]{Stevenson2014a}
{Stevenson} KB, {Bean} JL, {Madhusudhan} N {Harrington} J (2014{\natexlab{a}})
  {Deciphering the Atmospheric Composition of WASP-12b: A Comprehensive
  Analysis of its Dayside Emission}. \apj 791:36

\bibitem[{{Stevenson} et~al.(2014{\natexlab{b}}){Stevenson}, {D{\'e}sert},
  {Line}, {Bean}, {Fortney}, {Showman}, {Kataria}, {Kreidberg}, {McCullough},
  {Henry}, {Charbonneau}, {Burrows}, {Seager}, {Madhusudhan}, {Williamson}, and
  {Homeier}}]{Stevenson2014b}
{Stevenson} KB, {D{\'e}sert} JM, {Line} MR et~al. (2014{\natexlab{b}}) {Thermal
  structure of an exoplanet atmosphere from phase-resolved emission
  spectroscopy}. Science 346:838--841

\bibitem[{{Stevenson} et~al.(2017){Stevenson}, {Line}, {Bean}, {D{\'e}sert},
  {Fortney}, {Showman}, {Kataria}, {Kreidberg}, and {Feng}}]{Stevenson2017}
{Stevenson} KB, {Line} MR, {Bean} JL et~al. (2017) {Spitzer Phase Curve
  Constraints for WASP-43b at 3.6 and 4.5 {$\mu$}m}. \aj 153:68

\bibitem[{{Tsai} et~al.(2014){Tsai}, {Dobbs-Dixon}, and {Gu}}]{Tsai2014}
{Tsai} SM, {Dobbs-Dixon} I {Gu} PG (2014) {Three-dimensional Structures of
  Equatorial Waves and the Resulting Super-rotation in the Atmosphere of a
  Tidally Locked Hot Jupiter}. \apj 793:141

\bibitem[{{Visscher} and {Moses}(2011)}]{Visscher2011}
{Visscher} C {Moses} JI (2011) {Quenching of Carbon Monoxide and Methane in the
  Atmospheres of Cool Brown Dwarfs and Hot Jupiters}. \apj 738:72

\bibitem[{{Visscher} et~al.(2006){Visscher}, {Lodders}, and
  {Fegley}}]{Visscher2006}
{Visscher} C, {Lodders} K {Fegley} B Jr (2006) {Atmospheric Chemistry in Giant
  Planets, Brown Dwarfs, and Low-Mass Dwarf Stars. II. Sulfur and Phosphorus}.
  \apj 648:1181--1195

\bibitem[{{Webber} et~al.(2015){Webber}, {Lewis}, {Marley}, {Morley},
  {Fortney}, and {Cahoy}}]{Webber2015}
{Webber} MW, {Lewis} NK, {Marley} M et~al. (2015) {Effect of
  Longitude-dependent Cloud Coverage on Exoplanet Visible Wavelength
  Reflected-light Phase Curves}. \apj 804:94

\bibitem[{{Wong} et~al.(2015){Wong}, {Knutson}, {Lewis}, {Kataria}, {Burrows},
  {Fortney}, {Schwartz}, {Agol}, {Cowan}, {Deming}, {D{\'e}sert}, {Fulton},
  {Howard}, {Langton}, {Laughlin}, {Showman}, and {Todorov}}]{Wong2015}
{Wong} I, {Knutson} HA, {Lewis} NK et~al. (2015) {3.6 and 4.5 {$\mu$}m Phase
  Curves of the Highly Irradiated Eccentric Hot Jupiter WASP-14b}. \apj 811:122

\bibitem[{{Wong} et~al.(2016){Wong}, {Knutson}, {Kataria}, {Lewis}, {Burrows},
  {Fortney}, {Schwartz}, {Shporer}, {Agol}, {Cowan}, {Deming}, {D{\'e}sert},
  {Fulton}, {Howard}, {Langton}, {Laughlin}, {Showman}, and
  {Todorov}}]{Wong2016}
{Wong} I, {Knutson} HA, {Kataria} T et~al. (2016) {3.6 and 4.5 {$\mu$}m Spitzer
  Phase Curves of the Highly Irradiated Hot Jupiters WASP-19b and HAT-P-7b}.
  \apj 823:122

\bibitem[{{Yang} et~al.(2013){Yang}, {Cowan}, and {Abbot}}]{Yang2013}
{Yang} J, {Cowan} NB {Abbot} DS (2013) {Stabilizing Cloud Feedback Dramatically
  Expands the Habitable Zone of Tidally Locked Planets}. \apjl 771:L45

\bibitem[{{Zellem} et~al.(2014){Zellem}, {Lewis}, {Knutson}, {Griffith},
  {Showman}, {Fortney}, {Cowan}, {Agol}, {Burrows}, {Charbonneau}, {Deming},
  {Laughlin}, and {Langton}}]{Zellem2014}
{Zellem} RT, {Lewis} NK, {Knutson} HA et~al. (2014) {The 4.5 {$\mu$}m
  Full-orbit Phase Curve of the Hot Jupiter HD 209458b}. \apj 790:53

\bibitem[{{Zhang} and {Showman}(2017)}]{Zhang2017}
{Zhang} X {Showman} AP (2017) {Effects of Bulk Composition on the Atmospheric
  Dynamics on Close-in Exoplanets}. \apj 836:73

\end{thebibliography}

\end{document}